\documentclass[12pt]{article}
\makeatletter
\newcommand{\singlespacing}{\let\CS=\@currsize\renewcommand{\baselinestretch}{1}\tiny\CS}
\usepackage{epsfig}

\usepackage{graphicx}
\usepackage{verbatim}   
\usepackage{color}      
\usepackage{subfigure}  
\usepackage{amsfonts}
 \usepackage{caption}
\usepackage{array}
\usepackage{booktabs}
\usepackage{multirow}
\newcolumntype{d}{D{.}{.}{2.5}}
\newcolumntype{C}{>{\centering}p}

\begin{document}
\baselineskip=24pt
\parskip = 10pt
\def \qed {\hfill \vrule height7pt width 5pt depth 0pt}
\newcommand{\ve}[1]{\mbox{\boldmath$#1$}}
\newcommand{\IR}{\mbox{$I\!\!R$}}
\newcommand{\1}{\Rightarrow}
\newcommand{\bs}{\baselineskip}
\newcommand{\esp}{\end{sloppypar}}
\newcommand{\be}{\begin{equation}}
\newcommand{\ee}{\end{equation}}
\newcommand{\beanno}{\begin{eqnarray*}}
\newcommand{\inp}[2]{\left( {#1} ,\,{#2} \right)}
\newcommand{\eeanno}{\end{eqnarray*}}
\newcommand{\bea}{\begin{eqnarray}}
\newcommand{\eea}{\end{eqnarray}}
\newcommand{\ba}{\begin{array}}
\newcommand{\ea}{\end{array}}
\newcommand{\nno}{\nonumber}
\newcommand{\dou}{\partial}
\newcommand{\bc}{\begin{center}}
\newcommand{\ec}{\end{center}}
\newcommand{\2}{\subseteq}
\newcommand{\cl}{\centerline}
\newcommand{\ds}{\displaystyle}
\newcommand{\mr}{\mathbb{R}}
\newcommand{\mn}{\mathbb{N}}
\def\refhg{\hangindent=20pt\hangafter=1}
\def\refmark{\par\vskip 2.50mm\noindent\refhg}

\title{\sc Point and Interval Estimation of Weibull Parameters Based on Joint Progressively Censored Data}
\author{\sc Shuvashree Mondal \footnote{Department of Mathematics and Statistics, Indian Institute of
Technology Kanpur, Pin 208016, India.} and Debasis Kundu \footnote{Department of Mathematics and Statistics, Indian Institute of
Technology Kanpur, Pin 208016, India. Corresponding Author: E-mail: kundu@iitk.ac.in, Phone no. 91-512-2597141, Fax no. 91-512-2597500.}}

\date{}
\maketitle

\begin{abstract}

The analysis of progressively censored data has received considerable attention in the last few years.  In this paper we consider the joint progressive censoring scheme for two populations.  It is assumed that the lifetime distribution of the items from the two populations follow Weibull distribution with the same shape but different scale parameters.  Based on the joint progressive censoring scheme first we consider the maximum likelihood estimators of the unknown parameters whenever they exist.
We provide the Bayesian inferences of the unknown parameters under a fairly general priors on the shape and scale parameters.  
The Bayes estimators and the associated credible intervals cannot be obtained in closed form, and we propose to use the 
importance sampling technique to compute the same.  Further, we consider the problem when it is known apriori that the expected 
lifetime of one population is smaller than the other.  We provide the order restricted classical and Bayesian inferences of the unknown parameters.
Monte Carlo simulations are performed 
to observe the performances of the different estimators and the associated confidence and credible intervals.  One real data set has been analyzed for illustrative purpose.  

\end{abstract}



\newpage

\section{\sc Introduction}

In any life testing experiment it is a common practice to terminate the experiment before all specimens under observation fail.  In a type-I censoring scheme the test is terminated at a prefixed time point whereas in a type-II censoring scheme the experiment continues until a certain number of failures occurs.  In a practical scenario it might be necessary to remove some of the experimental units during the experiment.  Different progressive censoring schemes allow removal of experimental units during the experiment.  A progressive type-II censoring scheme can be briefly described as follows.  It is assumed that $n$ items are put on a test.  Suppose $k$, $R_1, \ldots, R_k$ are non-negative integers such that $n-k = R_1 + \ldots + R_k$. At the time of the first failure say $t_1$, $R_1$ units are chosen at random from the remaining $(n-1)$ items and they are removed from the experiment.  Then 
at the time of the second failure, say $t_2$, $R_2$ units are chosen at random from the remaining $n-2 - R_1$ units and they are removed.  The process continues, finally at the time of the $k$-th failure all the remaining $R_k$ items are removed from the experiment, and the experiment stops.

During the second world war, due to the demand of the highly reliable military equipment, engineers started laboratory investigation instead of long awaited field experiments.  Due to tremendous pressure on cost and time, various schemes were  introduced to reduce the cost of testing and multi-stage censoring was one of them. As it has been mentioned in Herd (1956) that in many laboratory evaluations, due to limited allocation of funds there are attempts to study the factors  contributing to either the reliability or the unreliability of the components or the whole systems under study, as well as to estimate the reliability of the items.  In some investigations in order to measure these auxiliary variables, some system needed to be disassembled or subject to measurement which are destructive in nature or might change the expected lifetime.  In certain evaluations, systems are removed from the main experimental setup in order to measure certain specific characteristics.  Most of the cases systems are subjected to interaction of human operator and to asses the effect of the human operator on the efficiency of the active system, sometimes it requires to withdraw few systems from the main experimental set up.  During the reliability study of production prototype systems, the addition of engineering modifications also require analyses on multistage censoring.

In Montanari and Cacciari( 1988) experimenters conducted testing to evaluate endurance of XLPE-insulated  cable to 
electrical and thermal stress along with the ageing mechanism.  Some specimens were removed from the test at selected times or at a time of breakdowns for the measurements of electrical, mechanical, chemical properties along with micro structural analyses in order to evaluate aging process.  These measurements are destructive in nature.  In this experiment the data obtained consist of failure time and censored time of specimens.
Ng et al. (2009) presented a clinical study where multistage censored data arise quite naturally.  In a study with plasma cell Myeloma at National Cancer Institute on 112 patients, few patients were dropping out at the end of certain intervals whose survival was ensured at that time but no further follow up was available.  In his thesis, Herd (1956) discussed estimation under multistage censoring scheme, and refer the scheme as "multi-censored samples".  Later on different multistage censoring schemes were referred as progressive censoring schemes.  Cohen (1963) studied the importance of progressive censoring scheme in reliability  experiment and Cohen (1966) discussed several cases where censored data occurred quite naturally.    

To be more precise, in favor of progressive censoring scheme the following points can be mentioned.  In practical scenario experimental units might get damaged due to some unrelated reasons than the normal failure mechanism.  To incorporate these information into inference study, we can rely on progressive censoring schemes.  Sometimes the multistage censoring is also intentionally done to use censored units from one experiment to another related experiment due to budgetary constraints.  

The progressive censoring scheme has received considerable attention in the literature.  Mann (1971) and Lemon (1975) studied on estimation for Weibull parameters under progressive censoring scheme.  Viveros and Balakrishnan (1994) provided interval estimation on progressively censored data.  Ng et al. (2004) studied on optimal progressive censoring plan when the underlined distribution is Weibull distribution whereas  Kundu (2008) provided Bayesian inference of the Weibull population under progressive censoring scheme.  Wang et al. (2010) studied the inference of certain life time distributions under progressive type-II right censored scheme.
Recently published book by 
Balakrishnan and Cramer (2014) provided an excellent overview of the different developments on different aspects of the progressive 
censoring scheme, which have taken place on this topic 
over the last 20 years.  Although, type-II progressive censoring scheme is the most popular one, several other progressive 
censoring schemes have also been introduced in the literature.  In this article we have restricted our attention to the 
type-II progressive censoring scheme, although most of our results can be extended for other progressive censoring schemes also.

Although, extensive work has been done on the progressive censoring scheme for one group, not much work has been done when 
two or more groups are present.  Rasouli and Balakrishnan (2010) introduced a joint progressive censoring (JPC) scheme, which 
can be used to compare the lifetime 
distributions of two products manufactured in different units under the same environmental conditions.   The JPC scheme
proposed by Rasouli and Balakrishnan (2010) can be briefly described as follows.  Suppose $m$ units of Product A (Group 1) and 
$n$ units of product B (Group 2) are put on a test simultaneously  at time zero.
It is assumed that $R_1, \ldots, R_k$ are $k$ non-negative integers such that $R_1 + \ldots + R_k = n+m-k$.  
At the time of the first failure, which may be either
from Group 1 or Group 2, $R_1$ items from the remaining $n+m-1$ remaining items have been selected at random, and they have been 
removed from the experiment.  These $R_1$ censored units consist of $S_1$ units from Group 1 and $R_1-S_1$ units from Group 2.  Here $S_1$ is random.  The time and group of the first failed item are recorded.  Similarly,
at the time of the second failure, $R_2$ items from the remaining $n+m-R_1 - 2$ items have been chosen at random and they have been removed 
from the experiment.  Among censored $R_2$ units random number of $S_2$ units come from Group 1.  The time and group of the second failed items are recorded.   The process continues till the $k$-th failure takes place, when all the remaining items are removed and the experiment ends.  

Based on the assumptions that the lifetime distributions of the 
experimental units of the two populations follow exponential distribution with different scale parameters, Rasouli and Balakrishnan (2010)
provided the exact distributions of the maximum likelihood estimators (MLEs) of the unknown parameters and suggested several confidence 
intervals.  Some of the related work on JPC scheme are by Shafay et al. (2014), Parsi and Bairamov (2009) and  Doostparast et al. (2013).  In most of these 
cases it has been assumed that the lifetime distributions of the items in two groups follow exponential distribution.

The exponential distribution has the constant hazard rate, which may not be very reasonable in a practical scenario.  Again when two similar kind of products are tested, it is quite expected to have some common parameters from the underlined distributions.
To justify these scenario, in this paper it is assumed that the lifetime 
distributions of the individual items of the two different groups follow Weibull distribution with the same shape parameter,
but different scale parameters.   

Our aim is to compare the lifetime distributions of the two populations.  The problem is a typical two-sample problem.  It can appear in a accelerated life testing problem, or in estimating the stress-strength 
parameter of a system.

First we consider the MLEs of the unknown parameters based on the data obtained from a JPC scheme as 
proposed by Rasouli and Balakrishnan (2010).  It has been shown that the MLEs exist under a very general condition and they
are unique.  The MLE of the common shape parameter can be obtained by solving one non-linear equation, and given the MLE of the 
common shape parameter, the MLEs of the scale parameters can be obtained in explicit forms.  Although, the performances of the MLEs 
are quite satisfactory, the associated confidence intervals are not very easy to obtain.  Hence, it seems the Bayesian inference is a 
natural choice in this case.  It may be treated as an extension of the work of Kundu (2008),  where the 
Bayesian inference of the unknown Weibull parameters for one sample problem was considered, and in this case the results have 
been generalized to two sample problem.  Clearly, the generalization is a non-trivial generalization.  Although, the whole development in this paper is for two groups, but the results can be easily generalized to more than two groups also.

For the Bayesian inference we need to assume some priors on the unknown parameters.
If the common shape parameter is known, the most convenient but a fairly general conjugate prior on the scale parameters can be the Beta-Gamma prior, as it was suggested by Pena and Gupta (1990).  In this case the explicit form of the Bayes estimates of the scale parameters can be obtained.  When the shape parameter is unknown, the conjugate prior does not exist.  In this case following the approach of Berger and Sun (1993) or Kundu (2008), it is assumed that the prior on the shape parameter has a support on $(0, \infty)$, and it has a log-concave density function.  It may be mentioned that many well known distribution functions for example log-normal, Weibull, gamma etc. may have log-concave probability density function.  Based on the prior distributions, the joint posterior density function can be obtained.  As expected the explicit expressions of the Bayes estimates cannot be obtained in explicit forms.  We propose to use importance sampling technique to compute the Bayes estimate of any function of the unknown parameters, and also to construct the associated highest posterior density (HPD) credible interval.  Monte Carlo simulations are performed to see the performances of the proposed method, and one data analysis has been performed for illustrative  purposes.  

In many practical situations it is known apriori that one population is better than the other in terms of the expected lifetime of the experimental
units. In accelerated life testing if units of one sample are put on higher stress than the units of other sample, it is quite expected to get shorter life time of the specimens under higher stress.  We incorporate this information by considering the order restricted classical and Bayesian inference of the unknown parameters based on joint progressively censored samples.  We obtain the MLEs of the unknown parameters based on the order restriction.  The construction of the confidence intervals 
of the unknown parameters can be obtained using bootstrap method.  For Bayesian inference, we propose a new order restricted Beta-Gamma prior.  In this case the explicit expressions of the Bayes estimates cannot
be obtained and we propose to use importance sampling technique to compute the Bayes estimates and the associated HPD credible intervals.  We 
re-analyze the data set based on the order restriction.

Rest of the paper is organized as follows.  In Section 2, we present the notations, preliminaries and the priors.  Maximum likelihood
estimators are presented in Section 3.  Posterior analysis are provided in Section 4.  In Section 5, we present the order restricted 
inferences of the unknown parameters.  In Section 6, we 
present the simulation results and data analysis.  Finally we conclude the paper in Section 7.

\section{\sc Notations, Model Assumptions and Priors}

\subsection{\sc Notations and Model Assumptions}
$$
\matrix{
\hbox{PDF}: & \hbox{Probability density function}   \cr
\hbox{HPD}: & \hbox{Highest posterior density}  \cr
\hbox{CDF}: & \hbox{Cumulative distribution function}  \cr
\hbox{MLE}: & \hbox{Maximum likelihood estimator}  \cr
\hbox{i.i.d.}: & \hbox{independent and identically distributed} \cr
F_1: & \hbox{CDF of the lifetime distribution of the items of Group 1}  \cr
F_2: & \hbox{CDF of the lifetime distribution of the items of Group 2}  \cr
T_1: & \hbox{Random variable with CDF $F_1$}  \cr  
T_2: & \hbox{Random variable with CDF $F_2$}  \cr 
k_1: & \hbox{Number of failures from Group 1}  \cr
k_2: & \hbox{Number of failures from Group 2}  \cr
k: & \hbox{Total number of failures:} \ \ k = k_1+k_2   \cr
\hbox{GA}(\alpha,\lambda): & \hbox{Gamma random variable with PDF:    } 
\ds (\lambda^{\alpha}/\Gamma(\alpha)) x^{\alpha-1}
e^{-\lambda x}; \ \  x > 0.  \cr
\hbox{WE}(\alpha,\lambda): & \hbox{Weibull random variable with PDF:   } 
\ds \alpha \lambda x^{\alpha-1} e^{-\lambda x^{\alpha}}; \ \  x > 0.  \cr
\hbox{Bin}(n,p): & \hbox{Binomial random variable with probability mass function:    }
{n\choose{i}} p^i (1-p)^{n-i}   \cr
\hbox{Beta}(a,b): & \hbox{Beta random variable with PDF:    } \ds 
(\Gamma(a+b)/\Gamma(a) \Gamma (b)) p^{a-1} (1-p)^{b-1}; \ \  0 < p < 1.   \cr
}
$$

Suppose $m$ and $n$ independent units are placed on a test with the corresponding lifetimes being identically distributed
with PDF $f_1(\cdot)$ and $f_2(\cdot)$, and CDF $F_1(\cdot)$ and $F_2(\cdot)$, respectively.  It is assumed that the lifetime
distribution of the items from Group 1 and Group 2, follow WE$(\alpha,\lambda_1)$ and WE$(\alpha,\lambda_2)$, respectively.
For a given $(R_1, \ldots, R_k)$, as described before, in a JPC scheme, the observations are as follows;
\be
\{(t_1, \delta_1, s_1), \ldots, (t_k, \delta_k, s_k)\}.   \label{data}
\ee
Here for $j = 1, \ldots, k$, $\delta_j$ = 1, if the failure at $t_j$ occurs from Group 1, otherwise $\delta_j$ = 0, and $s_j$ denotes the number 
of items from Group 1, which have been removed at the time $t_j$.  Therefore, the likelihood function can be written as
\be
L(data|\alpha, \lambda_1, \lambda_2) \propto 
\alpha^k \lambda_1^{k_1} \lambda_2^{k_2} \prod_{j=1}^k t_j^{\alpha-1} 
e^{-\lambda_1 U(\alpha)} e^{-\lambda_2 V(\alpha)}, \ \ \ \ \label{lh-2}
\ee
here $\ds k_1 = \sum_{j=1}^k \delta_j$ and $k_2 = k - k_1$, 
$$
C_1 = \{j; \delta_j = 1\}, \ \ \ C_2 = \{j; \delta_j = 0\}, 
$$
$$
U(\alpha) = \sum_{j=1}^k s_j t_j^{\alpha} + \sum_{C_1} t_j^{\alpha}, \ \ \
V(\alpha) = \sum_{j=1}^k w_j t_j^{\alpha} + \sum_{C_2} t_j^{\alpha},
$$
where $w_j = R_j - s_j$.

\subsection{\sc Prior Assumptions: Without Order Restriction}

The following prior assumptions are made on the common shape parameter $\alpha$ and on the scale parameters $\lambda_1$ and
$\lambda_2$, when there is no order restriction on $\lambda_1$ and $\lambda_2$.  If we denote $\lambda = \lambda_1 + \lambda_2$, then similarly as in Pena and Gupta (1990), it is assumed that
$\lambda \sim$ GA($a_0,b_0$), with $a_0 > 0$, $b_0 > 0$ and $p = \lambda_1/(\lambda_1+\lambda_2) \sim$ Beta$(a_1,a_2)$, with 
$a_1 > 0$, $a_2 > 0$, and they are independently distributed.  The joint PDF of $\lambda_1$ and $\lambda_2$ can be obtained
as follows
\bea
\pi_1(\lambda_1, \lambda_2|a_0,b_0,a_1,a_2) & = & \frac{\Gamma(a_1+a_2)}{\Gamma(a_0)} (b_0 \lambda)^{a_0-a_1-a_2} \times 
\frac{b_0^{a_1}}{\Gamma(a_1)} \lambda_1^{a_1-1} e^{-b_0 \lambda_1}    \nonumber  \\
&  & \times  \frac{b_0^{a_2}}{\Gamma(a_2)} \lambda_2^{a_2-1} e^{-b_0 \lambda_2}; \ \ \ 0 < \lambda_1, \ \lambda_2 < \infty.   
\label{prior-1}
\eea
It is known as the Beta-Gamma PDF, and it will be denoted by BG$(a_0,b_0,a_1,a_2)$.

The above Beta-Gamma prior is a very flexible prior on the scale parameters.  Depending on the values of the hyper-parameters,
the joint prior on $\lambda_1$ and $\lambda_2$ can take variety of shapes.  Moreover, for a given $\alpha$, it is a conjugate 
prior on $(\lambda_1, \lambda_2)$.  The correlation between $\lambda_1$ and $\lambda_2$ can be both positive and negative 
depending on the values of $a_0$, $a_1$ and $a_2$.  If $a_0  = a_1 + a_2$, $\lambda_1$ and $\lambda_2$  are independent.  If
$a_0 > a_1 + a_2$, then they are positively correlated, and for $a_0 < a_1 + a_2$, they are negatively correlated.  The 
following results will be useful for further development, and they can be established very easily.

\noindent {\sc Result 1:} If $(\lambda_1, \lambda_2) \sim$ BG$(a_0,b_0,a_1,a_2)$, then for $i = 1,2$,
\be
E(\lambda_i) = \frac{a_0 a_i}{b_0(a_1+a_2)} \ \ \ \hbox{and} \ \ \ 
V(\lambda_i) = \frac{a_0 a_i}{b_0(a_1+a_2)} \times \left \{ \frac{(a_i+1)(a_0+1)}{a_1+a_2+1} - \frac{a_0 a_i}{a_1 + a_2} 
\right \}.    \label{mean-var}
\ee
Moreover, the generation from a Beta-Gamma distribution is quite simple using the property that 
$(\lambda_1, \lambda_2) \sim$ BG$(a_0,b_0,a_1,a_2)$, if and only, $\lambda_1+\lambda_2$ has a gamma distribution and 
$\ds \frac{\lambda_1}{\lambda_1+\lambda_2}$ has a beta distribution and they are independently distributed, see for 
example Kundu and Pradhan (2011).

No specific form of prior has been assumed here on the common shape parameter $\alpha$.  Following the idea of 
Berger and Sun (1993), it is assumed that $\pi(\alpha)$, the prior on $\alpha$, has a support on the positive real line, and
it has a log-concave PDF.  Moreover, $\pi(\alpha)$ and $\pi(\lambda_1, \lambda_1|b_0, a_),a_1,a_2)$ are independently 
distributed.  It may be mentioned that many well known distribution has log-concave PDF.  For example, log-normal, gamma, Weibull 
have log-concave PDFs.  $\pi(\alpha)$ also has its hyper parameters.  We do not make it explicit here,  whenever it is needed, we will make it explicit.  

\subsection{\sc Prior Assumptions: Order Restricted}

If we have a order restriction on the scale parameters as $\lambda_1 < \lambda_2$, then we make the following prior assumption
on $\lambda_1$ and $\lambda_2$.
\bea
\pi_2(\lambda_1, \lambda_2|a_0,b_0,a_1,a_2) & = & \frac{\Gamma(a_1+a_2)}{\Gamma(a_0) \Gamma(a_1)\Gamma(a_2)} 
b_0^{a_0}  \lambda^{a_0-a_1-a_2} e^{-b_0 (\lambda_1+\lambda_2)}   \times  \nonumber  \\
&  & \left (\lambda_1^{a_1-1} \lambda_2^{a_2-1} + \lambda_1^{a_2-1} \lambda_2^{a_1-1} \right ); \ \ \  0 < \lambda_1 < \lambda_2 < \infty.   
\label{prior-2}
\eea
We will call it as the ordered Beta-Gamma PDF, and it will be denoted by OBG($a_0, b_0, a_1, a_2$).  Note that (\ref{prior-2})
is the PDF of the ordered random variable $(\lambda_{(1)}, \lambda_{(2)})$, where $(\lambda_{(1)}, \lambda_{(2)}) = 
(\lambda_1, \lambda_2)$ if $\lambda_1 < \lambda_2$, $(\lambda_{(1)}, \lambda_{(2)}) = 
(\lambda_2, \lambda_1)$ if $\lambda_2 < \lambda_1$ and $(\lambda_1, \lambda_2)$ follows $(\sim$) BG($a_0, b_0, a_1, a_2$).
It may be noted that the generation from a ordered Beta-Gamma distribution is also quite straight forward.  First we 
generate sample from a Beta-Gamma distribution and then by ordering them, we obtain a random sample from a 
ordered Beta-Gamma distribution.
We assume the same prior $\pi(\alpha)$ on $\alpha$ as before and $\alpha$ and $(\lambda_1, \lambda_2)$ are assumed to be 
independently distributed.

\section{\sc Maximum Likelihood Estimators}

Based on the observations described in ({\ref{data}), the log-likelihood function without the additive constant becomes:
\be
l(data|\alpha,\lambda_1,\lambda_2)  =  k \ln \alpha +  k_1 \ln \lambda_1 +  k_2 \ln \lambda_2 + (\alpha-1) 
\sum_{j=1}^k \ln t_j - \lambda_1 U(\alpha) - \lambda_2 V(\alpha),   \label{ll-1}
\ee
here $k_1$, $k_2$, $U(\alpha)$ and $V(\alpha)$ are same as defined before.
The following result provides the uniqueness of the MLEs of $\lambda_1$ and $\lambda_2$ for a given $\alpha$.  

\noindent {\sc Theorem 1:} If $k_1 > 0$ and $k_2 > 0$, then for a fixed $\alpha > 0$, $g_1(\lambda_1, \lambda_2) = 
\ds l(data|\alpha,\lambda_1,\lambda_2)$
is a unimodal function of $(\lambda_1, \lambda_2)$.

\noindent {\sc Proof:} Note that $\ds g_1(\lambda_1, \lambda_2)$ is a concave function as the Hessian matrix of $g_1(\lambda_1, \lambda_2)$ 
is a negative definite matrix.    Now the result follows because
for fixed $\lambda_1 (\lambda_2)$, $g_1(\lambda_1, \lambda_2)$ tends to $-\infty$, as $\lambda_2 (\lambda_1)$ tends to 0, or 
$\infty$.   \qed

For known $\alpha$, the MLEs of $\lambda_1$ and $\lambda_2$, say $\widehat{\lambda}_1(\alpha)$ and $\widehat{\lambda}_2(\alpha)$, 
respectively, can be obtained as follows:
\be
\widehat{\lambda}_1(\alpha) = \frac{k_1}{U(\alpha)} \ \ \ \ \hbox{and} \ \ \ 
\widehat{\lambda}_2(\alpha) = \frac{k_2}{V(\alpha)}.     \label{mle-rl}
\ee
When $\alpha$ is unknown, first the MLE of $\alpha$ can be obtained by maximizing the profile log-likelihood function 
of $\alpha$ without the additive constant, and that is
\be
p_1(\alpha)  =  l(data|\alpha, \widehat{\lambda}_1(\alpha),\widehat{\lambda}_2(\alpha))  =    
k \ln \alpha - k_1 \ln U(\alpha) - k_2 \ln V(\alpha) + (\alpha-1) \sum_{j=1}^k \ln t_j.
\ee
The following result will provide the existence and uniqueness of the MLE of $\alpha$.

\noindent {\sc Theorem 2:} If $k_1 > 0$ and $k_2 > 0$, $p_1(\alpha)$ is a unimodal function of $\alpha$.

\noindent {\sc Proof:} See in the appendix A.

Therefore, combining Theorem 1 and Theorem 2, it is immediately obtained that for $k_1 > 0$ and $k_2 > 0$, the MLEs of 
$\alpha$, $\lambda_1$ and $\lambda_2$ are unique.  It is quite simple to compute the MLE of $\alpha$ as $p_1(\alpha)$ is a unimodal 
function.  Use the bisection or Newton-Raphson method to compute the MLEs of 
$\alpha$, and once the MLE of $\alpha$ is obtained the MLEs of $\lambda_1$ and $\lambda_2$ can be obtained 
from (\ref{mle-rl}).  Although, the MLEs can be obtained quite efficiently in this case, the exact distribution of the MLEs
is not possible to obtain.  Hence, the construction of confidence intervals of the unknown parameters may not be very simple.
The Fisher information matrix may be used to construct the asymptotic confidence intervals of the unknown parameters and it 
is provided in Appendix B.  Alternatively, we propose to use the bootstrap method to construct the confidence intervals of the 
unknown parameters in this case.  Since the exact confidence intervals cannot be obtained, the Bayesian inference seems to be a natural choice in this case.

\section {\sc Bayes Estimates and Credible Intervals}

In this section we provide the Bayes estimates of the unknown parameters, and the corresponding credible sets based on JPC
scheme as described before.  We mainly assume the squared error 
loss function, although any other loss function can be easily incorporated. 
Now to compute the Bayes estimates of the unknown parameters, we need to assume some specific form of the prior  
distribution of $\alpha$, and it is assumed that $\pi(\alpha)$ has a Gamma($a, b)$ distribution.  Hence, 
the joint posterior density function of $\lambda_1, \lambda_2$ and $\alpha$ can be written as
\bea
\pi(\lambda_1, \lambda_2, \alpha| Data) & \propto & (\lambda_1+\lambda_2)^{a_0-a_1-a_2} \lambda_1^{a_1+k_1-1} \lambda_2^{a_2+k_2-1}
e^{-\lambda_1(b_0+U(\alpha))} e^{-\lambda_2(b_0+V(\alpha))}   \nonumber  \\ 
&  & \alpha^{k+a-1} e^{-b \alpha}   \prod_{i=1}^k t_i^{\alpha}.   \label{post}
\eea
We re-write (\ref{post}) in the following manner
\bea
\pi(\lambda_1, \lambda_2, \alpha| Data) & \propto & (\lambda_1+\lambda_2)^{a_0-a_1-a_2} \lambda_1^{a_1+k_1-1} \lambda_2^{a_2+k_2-1}
e^{-(\lambda_1+\lambda_2)(b_0+W(\alpha))}    \nonumber  \\ 
&  & \alpha^{k+a-1} e^{-\alpha(b - \sum_{i=1}^k \ln t_i)} \times e^{-\lambda_1(U(\alpha) - W(\alpha))} e^{-\lambda_2(V(\alpha) - W(\alpha)}.   \label{post-2}
\eea
Here $W(\alpha) = \min\{U(\alpha), V(\alpha)\}$.  The posterior density function 
of $\alpha$, $\lambda_1$ and $\lambda_2$ can be written as
\be
\pi(\alpha, \lambda_1, \lambda_2|Data) \propto \pi_1^*(\lambda_1,\lambda_2|data, \alpha) \times 
\pi_2^*(\alpha|data) \times g(\alpha,\lambda_1,\lambda_2|data).
\ee
Here $\pi_1^*(\lambda_1, \lambda_2|\alpha, Data)$ is the PDF of a BG$\ds (a_0+k_1+k_2,b_0+W(\alpha),a_1+k_1, a_2+k_2)$, and
\be
\pi_2^*(\alpha) \propto \frac{\alpha^{k+a-1} e^{-\alpha((b - \sum_{i=1}^k \ln t_i))}}{(b_0+W(\alpha))^{a_0+k}}, \ \ \ \    
g(\alpha, \lambda_1, \lambda_2|Data) = e^{-\lambda_1(U(\alpha) - W(\alpha))} e^{-\lambda_2(V(\alpha) - W(\alpha))}
\ee
Therefore, the Bayes estimate of $h(\alpha,\lambda_1,\lambda_2)$, any function of $\alpha$, $\lambda_1$, $\lambda_2$ 
with respect to squared error loss function is 
\be
E_{\pi(\alpha,\lambda_1,\lambda_2|Data)}(h(\alpha,\lambda_1,\lambda_2))  =  
\int_0^{\infty} \int_0^{\infty} \int_0^{\infty} h(\alpha,\lambda_1,\lambda_2)
\pi(\alpha, \lambda_1, \lambda_2|Data) d\alpha d \lambda_1 d \lambda_2  = \frac{K_1}{K_2},    \label{post-exp}
\ee
provided it exists.  Here
\be
K_1 = 
\int_0^{\infty} \int_0^{\infty} \int_0^{\infty} h(\alpha,\lambda_1,\lambda_2) \times 
\pi_1^*(\lambda_1,\lambda_2|Data, \alpha) \times 
\pi_2^*(\alpha|data) \times g(\alpha,\lambda_1,\lambda_2|Data) d\alpha d \lambda_1 d \lambda_2.    \label{exp-k1}
\ee
and
\be
K_2 = 
\int_0^{\infty} \int_0^{\infty} \int_0^{\infty}  
\pi_1^*(\lambda_1,\lambda_2|Data, \alpha) \times 
\pi_2^*(\alpha|data) \times g(\alpha,\lambda_1,\lambda_2|Data) d\alpha d \lambda_1 d \lambda_2.    \label{exp-k2}
\ee
It is clear that (\ref{post-exp}) cannot be obtained in closed form.  We may use Lindley's approximation to compute 
(\ref{post-exp}), but we may not be able to compute the associated credible interval using that.  We propose to use 
importance sampling technique to compute simulation consistent Bayes estimate and the associated credible interval.
The details will be explained later.  We use the following result for that purpose.

\noindent {\sc Theorem 3:} The density function $\pi_2^*(\alpha|Data)$ is log-concave.

\noindent {\sc Proof:} It can be obtained along the same line as in Theorem 1, the details are avoided.    \qed

Therefore, it is quite simple to generate samples from $l(\alpha|Data)$ using the method of Devroye (1984) or Kundu (2008), and
for a given $\alpha$, random samples from  $l(\lambda_1, \lambda_2|Data, \alpha)$ can be easily generated using 
the method of Kundu and Pradhan (2011).  The following algorithm can be used to compute the Bayes estimate and also the 
associated HPD credible interval of $h(\alpha,\lambda_1,\lambda_2)$.

\noindent {\sc Algorithm}

\begin{itemize}

\item Step 1: Generate $\alpha$ from $\pi^*_2(\alpha|data)$.

\item Step 2: For a given $\alpha$, generate $\lambda_1$ and $\lambda_2$ from $\pi^*_1(\lambda_1,\lambda_2|data,\alpha)$. 

\item Step 3: Repeat the procedure $N$ times to generate $(\alpha_1, \lambda_{11}, \lambda_{21}), \ldots, (\alpha_N, \lambda_{1N}, 
\lambda_{2N})$. 
 
\item Step 4: To obtain Bayes estimate of $h(\alpha, \lambda_1, \lambda_2)$, compute $(h_1,\ldots, h_N)$, where $h_i=h(\alpha_i, \lambda_{1i},  \lambda_{2i})$ as well as compute $(g_1, \ldots , g_N)$, where $g_i=g(\alpha_i, \lambda_{1i},  \lambda_{2i})$.
\item Step 5:  Bayes estimate of $h(\beta, \lambda_1, \lambda_2)$ can be approximated as $\frac{\sum_{i=1}^{N}g_ih_i}{\sum_{j=1}^{N}g_j} =\sum_{i=1}^{N}{v}_i h_i$ where ${v}_i=\frac{g_i}{\sum_{j=1}^{N}g_j}$.
\item Step 6: To compute $100(1-\beta)\%$ CRI of $h( \alpha, \lambda_1, \lambda_2)$ , arrange ${h_i}$ in ascending order to obtain $(h_{(1)},\ldots h_{(N)})$ and record the corresponding ${v_i}$ as $(v_{(1)}, \ldots, v_{(N)})$.  A $100(1-\gamma)\%$ CRI can be obtained as $(h_{(j_1)}, h_{(j_2)})$ where $j_1, j_2$ such that 

\bea
\label{eq}
j_1 < j_2, \quad  j_1 , j_2 \in  \{ 1, \ldots , N \} \quad  \mbox{and} \quad \sum_{i=j_1}^{j_2}v_i \leq 1-\beta <\sum_{i=j_1}^{j_2+1}v_i
\eea

The $100(1-\beta)\%$ highest posterior density (HPD) CRI can be obtained as $(h_{(j^*_1)}, h_{(j^*_2)})$, such that $h_{(j^*_2)} -h_{(j^*_1)}\leq  h_{(j_2)} - h_{(j_1)}$ and $j^*_1, j^*_2$ satisfying (\ref{eq}) for all $j_1, j_2$ satisfying (\ref{eq}).
  
\end{itemize}

\section{\sc Order Restricted Inference}

In this section we consider the inference on the unknown parameters under the restriction $\lambda_1 < \lambda_2$.  In many practical 
situations experimenter may have the information that one population has a smaller expected lifetime than the other.  In our case, this
leads to the above restriction.  Therefore, our problem can be stated as follows.  Based on the same set of assumptions as in Section 2.2
and with $\lambda_1 < \lambda_2$, the problem is to estimate the unknown parameters $\alpha$, $\lambda_1$ and $\lambda_2$ using the data (\ref{data}). 

\subsection{\sc Maximum Likelihood Estimators}

In this case also we proceed along the same line as before.  For a given  $\alpha$, the log-likelihood function (\ref{ll-1}) 
is concave as a function of $\lambda_1$ and $\lambda_2$ and it has a unique maximum.  The maximum value of the function (\ref{ll-1}) 
is obtained at the point ($\widehat{\lambda}_1(\alpha), \widehat{\lambda}_2(\alpha)$).  Clearly if $\widehat{\lambda}_1(\alpha) < 
\widehat{\lambda}_1(\alpha)$, then the order restricted MLEs of $\lambda_1$ and $\lambda_2$, say $\widetilde{\lambda}_1(\alpha) =
\widehat{\lambda}_1(\alpha)$ and $\widetilde{\lambda}_2(\alpha) = \widetilde{\lambda}_2(\alpha)$, respectively.  On the other hand
if $\widehat{\lambda}_1(\alpha) \ge \widehat{\lambda}_1(\alpha)$, then the maximum value of $g_1(\lambda_1, \lambda_2)$ will be on 
the line $\lambda_1 = \lambda_2$ under the order restriction $\lambda_1 < \lambda_2$.  Therefore, in this case
$$
\widetilde{\lambda}_1(\alpha) = \widetilde{\lambda}_2(\alpha) = \hbox{arg max } g_1(\lambda, \lambda).
$$
Hence for a given $\alpha$, the order restricted MLEs of $\lambda_1$ and $\lambda_2$ become
\be
(\widetilde{\lambda}_1(\alpha), \widetilde{\lambda}_2(\alpha)) = 
\left \{ \matrix{(\widehat{\lambda}_1(\alpha), \widehat{\lambda}_2(\alpha)) & \hbox{if} & \widehat{\lambda}_1(\alpha) < 
\widehat{\lambda}_2(\alpha)  \cr
\left (\frac{k}{\sum_{j=1}^k(R_j+1)t_j^{\alpha}}, \frac{k}{\sum_{j=1}^k(R_j+1)t_j^{\alpha}} \right )
& \hbox{if} & \widehat{\lambda}_1(\alpha) \ge \widehat{\lambda}_2(\alpha).  \cr   \label{rl1rl2}
}  \right .
\ee
The MLE of $\alpha$, say $\widetilde{\alpha}$ can be obtained by maximizing $p_2(\alpha) = l(data|\alpha, \widetilde{\lambda}_1(\alpha), \widetilde{\lambda}_2(\alpha))$ with respect to $\alpha$.  The following result provides the existence and uniqueness of the MLE of 
$\alpha$.

\noindent {\sc Theorem 4:} If $k_1 > 0$ and $k_2 > 0$, $p_2(\alpha)$ is a unimodal function of $\alpha$.

\noindent {\sc Proof:} The result follows along the same line as in Theorem 2 by observing the fact that $p_2(\alpha)$ is log-concave
in both the region, and $p_2(\alpha)$ is a continuous function of $\alpha$.  \qed

Once the MLE of $\alpha$ is obtained, the MLEs of $\lambda_1$ and $\lambda_2$ can be obtained from (\ref{rl1rl2}) explicitly.  We 
propose to use bootstrap method to construct the confidence intervals of the unknown parameters.

\subsection{\sc Bayes Estimates and Credible Intervals}

In this section we will provide the order restricted Bayesian inference of the unknown parameters based on the prior assumptions as
provided in Section 2.4.  Similarly as before, for specific implementation we assume that $\pi(\alpha)$ has a Gamma$(a,b)$ distribution.
The joint posterior density function of $\alpha$, $\lambda_1$ and $\lambda_2$ for 
$\alpha > 0, 0 < \lambda_1 < \lambda_2$, can be written as
\bea
\pi(\lambda_1, \lambda_2, \alpha| Data) & \propto & (\lambda_1+\lambda_2)^{a_0-a_1-a_2} \left ( \lambda_1^{a_1-1} \lambda_2^{a_2-1} 
+ \lambda_1^{a_2-1} \lambda_2^{a_1-1} \right ) \lambda_1^{k_1} \lambda_2^{k_2}
  \nonumber  \\ 
&  & e^{-\lambda_1(b_0+U(\alpha))} e^{-\lambda_2(b_0+V(\alpha))}  \alpha^{k+a-1} e^{-b \alpha}   \prod_{i=1}^k t_i^{\alpha}.    \label{post-3}
\eea
We re-write (\ref{post-3}) as follows
\bea
\pi(\lambda_1, \lambda_2, \alpha| Data) & \propto & (\lambda_1+\lambda_2)^{a_0-a_1-a_2} \left ( \lambda_1^{a_1+J-1} \lambda_2^{a_2+J-1} 
+ \lambda_1^{a_2+J-1} \lambda_2^{a_1+J-1} \right ) e^{-(\lambda_1+\lambda_2)(b_0+W(\alpha))}
  \nonumber  \\ 
&  & \lambda_1^{k_1-J} \lambda_2^{k_2-J} e^{-\lambda_1(U(\alpha)-W(\alpha))} e^{-\lambda_2(V(\alpha)-W(\alpha))}  
\alpha^{k+a-1} e^{-\alpha(b - \sum_{i=1}^k \ln t_i)},     \label{post-4}
\eea
here $J = \min\{k_1, k_2\}$ and $W(\alpha)$ is same as defined before.

The posterior density function of $\alpha$, $\lambda_1$ and $\lambda_2$ in this case can be written as
\be
\pi(\alpha, \lambda_1, \lambda_2|Data) \propto \pi_1^*(\lambda_1,\lambda_2|data, \alpha) \times 
\pi_2^*(\alpha|data) \times g(\alpha,\lambda_1,\lambda_2|data).
\ee
Here $\pi_1^*(\lambda_1, \lambda_2|\alpha, Data)$ is the PDF of a OBG$\ds (a_0+2J,b_0+W(\alpha),a_1+J, a_2+J)$, and
\be
\pi_2^*(\alpha) \propto \frac{\alpha^{k+a-1} e^{-\alpha((b - \sum_{i=1}^k \ln t_i))}}{(b_0+W(\alpha))^{a_0+2J}}
\ee
and
\be
g(\alpha, \lambda_1, \lambda_2|Data) = \lambda_1^{k_1-J} \lambda_2^{k_2-J} e^{-\lambda_1(U(\alpha) - W(\alpha))} e^{-\lambda_2(V(\alpha) - W(\alpha))}.
\ee
Since $\pi_2^*(\alpha|Data)$ is a log-concave function, and the generation from a $\pi_1^*(\lambda_1, \lambda_2|\alpha)$ can be 
performed quite conveniently, we can use the same importance sampling technique as in Section 4, to compute the Bayes estimate
and the associated HPD credible interval of any function of $\alpha$, $\lambda_1$ and $\lambda_2$.

\section{\sc Numerical Experiments and Data Analysis}

\subsection{\sc Numerical Experiments}

In this section we have performed some simulation experiments to see the effectiveness of the proposed methods and also to 
observe how the ordered restricted inference behaves in practice.  We have considered $m$ = 20, $n$ = 25, $\lambda_1$ = 1.0, 
$\lambda_2$ = 0.5.  We have taken different effective sample sizes, $k$ = 20, 25, different censoring schemes, and different 
$\alpha$ values, $\alpha$ = 1 and 2.  

In each case we obtain the MLEs of the unknown parameters.  We compute the average estimates (AE) and the mean squared errors (MSE) based
on 10,000 replications.  The results are reported in Table \ref{table-1} and Table \ref{table-2}.  In each case we also compute 
90\% symmetric percentile bootstrap confidence interval based on 500 bootstrap samples.  We repeat the process 1000 times, and 
obtain the average lengths (AL) of the confidence intervals and their coverage percentages (CP).  The results are reported in Table \ref{table-3} and Table \ref{table-4}.

We further compute the Bayes 
estimates and the associated 90\% credible intervals based on both informative priors (IP) and non-informative priors (NIP).  
In case of non-informative 
priors it is assumed that $b_0 = a_0 = a_1 = a_2 = a = b$ = 0.  For informative priors, when $\alpha$ = 1,
then $b_0$ = 1, $\ds a_0 = \frac{3}{2} b_0$, $a_1$ = 2, $a_2$ = 4, $a$ = 2, $b$ = 2, and when $\alpha$ = 2, 
$b_0$ = 1, $\ds a_0 = \frac{3}{2} b_0$, $a_1$ = 2, $a_2$ = 4, $a$ = 4, $b$ = 2.  We compute the Bayes estimates and the associated credible intervals based on 1000 samples.  We report the average Bayes
estimates and the corresponding MSEs in Table \ref{table-1} and Table \ref{table-2}.  The average lengths of the credible intervals
and the associated coverage percentages are reported in Table \ref{table-3} and Table \ref{table-4}.  We use the following notation for a particular progressive censoring scheme.  For example $k=6$ and $R = (4, 0_{(5)})$ means $R_1$ = 4, $R_2 = R_3 = R_4 = R_5 = R_6=0$.

\begin{table}[h]
\caption{ $m=20,n=22,\alpha=1,\lambda_1=0.5, \lambda_2=1$}
\label{table-1}
\begin{center}
\scalebox{.7}{
\begin{tabular}{llllllll}
\toprule
\multicolumn{1}{c}{Censoring scheme} & \multicolumn{1}{c}{Parameter}
&\multicolumn{2}{c}{MLE} & \multicolumn{2}{l}{Bayes IP}  & \multicolumn{2}{l}{Bayes NIP}\\
&&\multicolumn{1}{c}{AE} & \multicolumn{1}{c}{MSE}
&\multicolumn{1}{c}{AE}  &\multicolumn{1}{l}{MSE} & \multicolumn{1}{l}{AE} &\multicolumn{1}{l}{MSE} \\
\midrule
k=20,R=(7,0$_{(18)}$,15) & $\alpha$ & 1.097 &0.063 & 1.075 &0.050 &1.095& 0.064\\
 & $\lambda_1$ &0.554&0.057&0.536 &0.035 &0.543 &0.047\\
 & $\lambda_2$ &1.102&0.147&1.066&0.086&1.086&0.123\\
 \midrule
 k=20,R=(0$_{(9)}$ ,7,0$_{(9)}$,15) &$\alpha$ & 1.101 &0.067&1.062&0.048 &1.066 &0.055\\
 &$\lambda_1$ & 0.562&0.062&0.535&0.031&0.557&0.057\\
 & $\lambda_2$ &1.123 &0.185&1.064&0.077&1.081&0.127\\
 \midrule
 k=20,R=(0$_{(18)}$,7,15) & $\alpha$ &1.105&0.072&1.068&0.052&1.064&0.060\\
 & $\lambda_1$ &0.569&0.074&0.548&0.035&0.551&0.052\\
 & $\lambda_2$ &1.126&0.191&1.077&0.092&1.085&0.117\\
 \midrule
 k=25,R=(7,0$_{(23)}$,10) &$\alpha$ & 1.077 & 0.044 & 1.063&0.043&1.064&0.040\\
   & $\lambda_1$ & 0.532&0.036&0.523 & 0.025&0.532& 0.036\\
   & $\lambda_2$ & 1.062&0.095 &1.042 & 0.068&1.044&0.085\\
   \midrule
 k=25,R=(0$_{(11)}$,7,0$_{(12)}$,10) & $\alpha$ & 1.078&0.044& 1.062&0.036&1.075&0.044\\
 & $\lambda_1$ & 0.537 & 0.038 & 0.524&0.025&0.530&0.034\\
 & $\lambda_2$ & 1.071&0.099& 1.053&0.069 &1.064&0.101\\
 \midrule
 k=25,R=(0$_{(23)}$,7,10)& $\alpha$ & 1.080 & 0.049 & 1.073& 0.043&1.079&0.047\\
 & $\lambda_1$ & 0.540& 0.040 & 0.529&0.026&0.535&0.035\\
 & $\lambda_2$ & 1.071 & 0.099&1.059&0.075&1.069&0.098\\
 \bottomrule

\end{tabular} 
}
\end{center}

\end{table}

\begin{table}[h]
\caption{ $m=20,n=22,\alpha=2,\lambda_1=0.5, \lambda_2=1$}
\label{table-2}
\begin{center}
\scalebox{.7}{
\begin{tabular}{llllllll}
\toprule
\multicolumn{1}{c}{Censoring scheme} & \multicolumn{1}{c}{Parameter}
&\multicolumn{2}{c}{MLE} & \multicolumn{2}{l}{Bayes IP}  & \multicolumn{2}{l}{Bayes NIP}\\
&&\multicolumn{1}{c}{AE} & \multicolumn{1}{c}{MSE}
&\multicolumn{1}{c}{AE}  &\multicolumn{1}{l}{MSE} & \multicolumn{1}{l}{AE} &\multicolumn{1}{l}{MSE} \\
\midrule
k=20,R=(7,0$_{(18)}$,15) & $\alpha$ &2.191 &0.252 & 2.127 &0.157&2.185&0.243\\
 & $\lambda_1$ &0.555&0.057&0.545&0.035&0.550&0.049\\
 & $\lambda_2$ &1.097&0.143&1.055&0.078&1.104&0.131\\
 \midrule
 k=20,R=(0$_{(9)}$ ,7,0$_{(9)}$,15) &$\alpha$ &2.209&0.264&2.108&0.145&2.147&0.228\\
 &$\lambda_1$ &0.563&0.063&0.536&0.034&0.556&0.049\\
 & $\lambda_2$ &1.118&0.165&1.097&0.099&1.081&0.126\\
 \midrule
 k=20,R=(0$_{(18)}$,7,15) & $\alpha$ &2.207&0.273&2.120&0.157&2.160&0.251\\
 & $\lambda_1$ &0.561&0.067&0.548&0.035&0.541&.046\\
 & $\lambda_2$ &1.122&0.183&1.064&0.082&1.112&0.148\\
 \midrule
 k=25,R=(7,0$_{(23)}$,10) & $\alpha$ & 2.149 & 0.169 &2.133&0.151&2.139&0.171\\
 & $\lambda_1$ & 0.534 & 0.037 &0.523&0.025&0.524&0.034\\
 & $\lambda_2$ & 1.065&0.101&1.056&0.073&1.061&0.094\\
 \midrule
 k=25,R=(0$_{(11)}$,7,0$_{(12)}$,10) & $\alpha$ & 2.150& 0.176 &2.127&0.135&2.146&0.192\\
 & $\lambda_1$ & 0.5380& 0.036 &0.523&0.023&0.541&0.035\\
 & $\lambda_2$ & 1.066&0.102&1.045&0.068&1.047&0.088\\
 \midrule
 k=25,R=(R=(0$_{(23)}$,7,10) & $\alpha$ & 2.156 &0.193 &2.122&0.144&2.132&0.191\\
  & $\lambda_1$ &0.537 & 0.038 &0.527&0.025&0.545&0.039\\
  & $\lambda_2$ &1.071&0.101&1.048&0.062&1.049&0.093 \\
 \bottomrule
\end{tabular} 
}
\end{center}
\end{table}

\begin{table}[h]
\caption{ $m=20,n=22,\alpha=1,\lambda_1=0.5, \lambda_2=1$}
\label{table-3}
\begin{center}
\scalebox{.7}{
\begin{tabular}{llllllll}
\toprule
\multicolumn{1}{c}{Censoring scheme} & \multicolumn{1}{c}{Parameter} & \multicolumn{2}{l}{90\% HPD CRI IP}  & \multicolumn{2}{l}{90\% HPD CRI NIP} &\multicolumn{2}{l}{90\% Bootstrap CI}\\
&&\multicolumn{1}{c}{AL} & \multicolumn{1}{l}{CP}
&\multicolumn{1}{c}{AL}  &\multicolumn{1}{l}{CP} &\multicolumn{1}{c}{AL}  &\multicolumn{1}{l}{CP}\\
\midrule
k=20,R=(7,0$_{(18)}$,15) & $\alpha$ & 0.621 & 86.6\% &0.663&82.8\% &0.804 & 82.2\%\\
 & $\lambda_1$ &0.560&90.0\%&0.627&86.8\% & 0.814 & 87.0\%\\
 & $\lambda_2$&0.894&89.6\%&1.018&87.6\% &1.358&83.8\% \\
 \midrule
 k=20,R=(0$_{(9)}$ ,7,0$_{(9)}$,15) &$\alpha$ &0.603&84.6\%&0.624&85.0\% & 0.818&83.0\%\\
 &$\lambda_1$ &0.565&87.6\%&0.635&87.2\% &0.907&85.6\% \\
 & $\lambda_2$ &0.900&91.6\%&0.977&89.2\% &1.454&82.8\%\\
 \midrule
 k=20,R=(0$_{(18)}$,7,15) & $\alpha$ &0.615&87.0\%&0.624&82.0\% & 0.854 & 83.6\%\\
 & $\lambda_1$ &0.582&92.0\%&0.649&87.2\% & 0.900&87.2\%\\
 & $\lambda_2$ &0.903&91.4\%&0.981&89.4\%& 1.463 &86.0\%\\
 \bottomrule
\end{tabular} 
}
\end{center}
\end{table}

\begin{table}[h]
\caption{ $m=20,n=22,\alpha=2,\lambda_1=0.5, \lambda_2=1$}
\label{table-4}
\begin{center}
\scalebox{.7}{
\begin{tabular}{llllllll}
\toprule
\multicolumn{1}{c}{Censoring scheme} & \multicolumn{1}{c}{Parameter} & \multicolumn{2}{l}{90\% HPD CRI IP}  & \multicolumn{2}{l}{90\% HPD CRI NIP} &\multicolumn{2}{l}{90\% Bootstrap CI}\\
&&\multicolumn{1}{c}{AL} & \multicolumn{1}{l}{CP}
&\multicolumn{1}{c}{AL}  &\multicolumn{1}{l}{CP} &\multicolumn{1}{c}{AL} &\multicolumn{1}{l}{CP}\\
\midrule
k=20,R=(7,0$_{(18)}$,15) & $\alpha$ & 1.198 & 90.8\% &1.309&87.4\% &1.612&80.2\%\\
 & $\lambda_1$ &0.562&88.2\%&0.639&85.0\%&0.827&87.2\%\\
 & $\lambda_2$&0.897&90.8\%&0.993&89.6\% &1.293&87.8\%\\
 \midrule
 k=20,R=(0$_{(9)}$ ,7,0$_{(9)}$,15) &$\alpha$ &1.166&89.4\%&1.209&82.2\% &1.690&77.2\%\\
 &$\lambda_1$ &0.569&91.4\%&0.614&85.0\% &0.903&88.4\%\\
 & $\lambda_2$ &0.910&90.4\%&1.000&88.4\% &1.492&85.6\%\\
 \midrule
 k=20,R=(0$_{(18)}$,7,15) & $\alpha$ &1.171&89.0\%&1.258&81.8\% &1.758&81.0\%\\
 & $\lambda_1$ &0.572&90.6\%&0.652&85.0\% &0.902&87.8\%\\
 & $\lambda_2$ &0.907&91.8\%&0.988&89.0\%&1.468&84.4\%\\
 \bottomrule
\end{tabular} 
}
\end{center}
\end{table}

From Table \ref{table-1} and Table \ref{table-2}, it is clear that as the 
effective sample size increases in all the cases the average biases and the MSEs decrease.  The performances of the Bayes estimators with respect to the NIPs  are slightly better than the MLEs both in terms of biases and MSEs in most of the cases investigated here.  Also the Bayes estimators
with respect to IPS are performing better than the Bayes estimators with respect to NIPs in terms of the biases and MSEs, as expected.  
In Table \ref{table-3} and Table \ref{table-4} we report the performances of the percentile bootstrap confidence intervals and the Bayes credible intervals.  In most of the cases the coverage percentages are very close to the nominal values.  The average credible lengths based on the non-informative priors are larger than the informative priors, but they are slightly lower than the average bootstrap confidence intervals.

Therefore, it is clear in this case that the Bayesian inference is preferable than the classical inference.
If we have some prior information on the unknown parameters then we should use the Bayes estimates and the 
associated credible intervals with respect to the informative priors, otherwise we should use the non-informative priors.

Further, we perform the order restricted inference of the unknown parameters.  In this case we have taken the same set of 
parameter values, the sample sizes and the censoring schemes mainly for comparison purposes.  For the Bayesian inference we have considered the same set of 
hyper parameters also.  The average estimates and the associated MSEs are reported in Table \ref{table-5} and Table \ref{table-6}.
The confidence and credible intervals are reported in Table \ref{table-7} and Table \ref{table-8}.  

From Tables \ref{table-5} and 
\ref{table-6}, it is clear that the performances of the Bayes estimators based on the non-informative priors behave slightly better than the MLEs in terms of MSEs
particularly for the scale parameters.  In this case also the performances of the Bayes estimators with respect to the informative
priors perform much better than the non-informative priors, as expected.  From Tables \ref{table-7} and \ref{table-8}, it is observed 
that coverage percentages of the bootstrap confidence intervals are very similar to the Bayes estimators based on the non-informative
priors.  The coverage percentages of the bootstrap confidence intervals are slightly closer to the nominal value than the Bayes
estimator with respect to the non-informative priors in most of the cases considered here.

From above discussion we can conclude that when the order restriction is imposed on the scale parameters, and if we have some prior information 
about the unknown parameters, then the Bayesian inference with respect to the informative priors is preferable, otherwise non-informative
priors may be used.

\begin{table}[h]
\caption{ $m=20,n=22,\alpha=1,\lambda_1=0.5, \lambda_2=1$ \\(With  Order Restriction)}
\label{table-5}
\begin{center}
\scalebox{.7}{
\begin{tabular}{llllllll}
\toprule
\multicolumn{1}{c}{Censoring scheme} & \multicolumn{1}{c}{Parameter}
&\multicolumn{2}{c}{MLE} & \multicolumn{2}{l}{Bayes inf-prior}  & \multicolumn{2}{l}{Bayes noninf-prior}\\
&&\multicolumn{1}{c}{AE} & \multicolumn{1}{c}{MSE}
&\multicolumn{1}{c}{AE}  &\multicolumn{1}{l}{MSE} & \multicolumn{1}{l}{AE} &\multicolumn{1}{l}{MSE} \\
\midrule
k=20,R=(7,0$_{(18)}$,15) & $\alpha$ &1.093& 0.063& 1.080&0.051&1.078&0.056\\
 & $\lambda_1$ &0.551&0.052&0.526&0.024&0.511&0.031\\
 & $\lambda_2$ &1.113&0.155&1.041&0.067&1.103&0.110\\
 \midrule
 k=20,R=(0$_{(9)}$ ,7,0$_{(9)}$,15) &$\alpha$ &1.102&0.064&1.057&0.055&1.077&0.057\\
 &$\lambda_1$ &0.553&0.058&0.527&0.026&0.506&0.034\\
 & $\lambda_2$ &1.124&0.178&1.051&0.076&1.095&0.109\\
 \midrule
 k=20,R=(0$_{(18)}$,7,15) & $\alpha$ &1.102&0.072&1.078&0.054&1.081&0.066\\
 & $\lambda_1$ &0.560&0.062&0.535&0.028&0.504&0.028\\
 & $\lambda_2$ &1.129&0.184&1.057&0.080&1.085&0.102\\
 \midrule
 k=25,R=(7,0$_{(23)}$,10) &$\alpha$ &1.075&0.043&1.081&0.052&1.066&0.045\\
   & $\lambda_1$ &0.529&0.033&0.523&0.020&0.508&0.022\\
   & $\lambda_2$ &1.068&0.986&1.032&0.062&1.056&0.081\\
   \midrule
 k=25,R=(0$_{(11)}$,7,0$_{(12)}$,10) & $\alpha$ &1.079&0.045&1.053&0.035&1.079&0.048\\
 & $\lambda_1$ &0.535&0.034&0.529&0.022&0.501&0.022\\
 & $\lambda_2$ &1.072&0.099&1.034&0.058&1.073&0.080\\
 \midrule
 k=25,R=(0$_{(23)}$,7,10)& $\alpha$ &1.082&0.050&1.061&0.042&1.070&0.046\\
 & $\lambda_1$ &0.536&0.036&0.526&0.021&0.508&0.023\\
 & $\lambda_2$ &1.075&0.101&1.029&0.058&1.065&0.075\\
 \bottomrule
\end{tabular} 
}
\end{center}
\end{table}

\begin{table}[h]
\caption{ $m=20,n=22,\alpha=2,\lambda_1=0.5, \lambda_2=1$\\ (With  Order Restriction)}
\label{table-6}
\begin{center}
\scalebox{.7}{
\begin{tabular}{llllllll}
\toprule
\multicolumn{1}{c}{Censoring scheme} & \multicolumn{1}{c}{Parameter}
&\multicolumn{2}{c}{MLE} & \multicolumn{2}{l}{Bayes inf-prior}  & \multicolumn{2}{l}{Bayes noninf-prior}\\
&&\multicolumn{1}{c}{AE} & \multicolumn{1}{c}{MSE}
&\multicolumn{1}{c}{AE}  &\multicolumn{1}{l}{MSE} & \multicolumn{1}{l}{AE} &\multicolumn{1}{l}{MSE} \\
\midrule
k=20,R=(7,0$_{(18)}$,15) & $\alpha$ &2.185&0.247&2.123&0.160&2.161&0.239\\
 & $\lambda_1$ &0.548&0.051&0.528&0.028&0.505&0.033\\
 & $\lambda_2$ &1.107&0.147&1.056&0.075&1.091&0.110\\
 \midrule
 k=20,R=(0$_{(9)}$ ,7,0$_{(9)}$,15) &$\alpha$ &2.197&0.266&2.108&0.153&2.146&0.219\\
 &$\lambda_1$ &0.555&0.057&0.524&0.025&0.508&0.032\\
 & $\lambda_2$ &1.119&0.167&1.046&0.074&1.092&0.113\\
 \midrule
 k=20,R=(0$_{(18)}$,7,15) & $\alpha$ &2.198&0.271&2.111&0.163&2.145&0.249\\
 & $\lambda_1$ &0.558&0.058&0.525&0.025&0.521&0.034\\
 & $\lambda_2$ &1.129&0.171&1.047&0.069&1.089&0.106\\
 \midrule
 k=25,R=(7,0$_{(23)}$,10) & $\alpha$ &2.149&0.175&2.127&0.171&2.121&0.157\\
 & $\lambda_1$ &0.527&0.033&0.517&0.021&0.508&0.023\\
 & $\lambda_2$ &1.061&0.094&1.034&0.063&1.072&0.082\\
 \midrule
 k=25,R=(0$_{(11)}$,7,0$_{(12)}$,10) & $\alpha$ &2.149&0.176&2.115&0.126&2.102&0.150\\
 & $\lambda_1$ &0.536&0.033&0.514&0.020&0.502&0.020\\
 & $\lambda_2$ &1.073&0.101&1.024&0.063&1.058&0.073\\
 \midrule
 k=25,R=(R=(0$_{(23)}$,7,10) & $\alpha$ &2.169&0.201&2.126&0.144&2.154&0.180\\
  & $\lambda_1$ &0.540&0.039&0.530&0.022&0.502&0.023\\
  & $\lambda_2$ &1.079&0.102&1.039&0.060&1.052&0.072\\
 \bottomrule
\end{tabular} 
}
\end{center}
\end{table}

\begin{table}[h]
\caption{ $m=20,n=22,\alpha=1,\lambda_1=0.5, \lambda_2=1$\\ (With Order Restriction)}
\label{table-7}
\begin{center}
\scalebox{.7}{
\begin{tabular}{llllllll}
\toprule
\multicolumn{1}{c}{Censoring scheme} & \multicolumn{1}{c}{Parameter} & \multicolumn{2}{l}{90\% HPD CRI inf-prior}  & \multicolumn{2}{l}{90\% HPD CRI noninf-prior}& \multicolumn{2}{l}{90\% Bootstrap CI}\\
&&\multicolumn{1}{c}{AL} & \multicolumn{1}{l}{CP}
&\multicolumn{1}{c}{AL}  &\multicolumn{1}{l}{CP}&\multicolumn{1}{c}{AL}  &\multicolumn{1}{l}{CP}\\
\midrule
k=20,R=(7,0$_{(18)}$,15) & $\alpha$ &0.600&83.8\% &0.624&82.2\%&0.801&80.8\%\\
 & $\lambda_1$ &0.502&88.6\%&0.515&86.2\%&0.740&89.2\%\\
 & $\lambda_2$&0.711&83.4\%&0.722&81.4\%&1.312&84.0\% \\
 \midrule
 k=20,R=(0$_{(9)}$ ,7,0$_{(9)}$,15) &$\alpha$ &0.587&83.6\%&0.612&82.4\% &0.813&81.4\%\\
 &$\lambda_1$ &0.515&88.5\%&0.537&87.2\% &0.763&89.8\%\\
 & $\lambda_2$ &0.685&80.0\%&0.731&81.4\%&1.368&86.6\%\\
 \midrule
 k=20,R=(0$_{(18)}$,7,15) & $\alpha$ &0.591&81.4\%&0.630&81.0\%&0.865&83.0\%\\
 & $\lambda_1$ &0.517&91.4\%&0.534&82.4\%&0.829&89.8\%\\
 & $\lambda_2$ &0.683&82.6\%&0.728&77.0\%&1.433&85.8\%\\
 \bottomrule
\end{tabular} 
}
\end{center}
\end{table}

\begin{table}[h]
\caption{ $m=20,n=22,\alpha=2,\lambda_1=0.5, \lambda_2=1$\\ (With Order Restriction)}
\label{table-8}
\begin{center}
\scalebox{.7}{
\begin{tabular}{llllllll}
\toprule
\multicolumn{1}{c}{Censoring scheme} & \multicolumn{1}{c}{Parameter} & \multicolumn{2}{l}{90\% HPD CRI inf-prior}  & \multicolumn{2}{l}{90\% HPD CRI noninf-prior}& \multicolumn{2}{l}{90\% Bootstrap CI}\\
&&\multicolumn{1}{c}{AL} & \multicolumn{1}{l}{CP}
&\multicolumn{1}{c}{AL}  &\multicolumn{1}{l}{CP}&\multicolumn{1}{c}{AL} &\multicolumn{1}{l}{CP}\\
\midrule
k=20,R=(7,0$_{(18)}$,15) & $\alpha$ &1.141&87.8\% &1.280&84.6\%&1.618&80.0\%\\
 & $\lambda_1$ &0.499&89.2\%&0.517&86.8\% &0.743&88.8\%\\
 & $\lambda_2$&0.724&83.4\%&0.746&79.8\% &1.314&85.4\%\\
 \midrule
 k=20,R=(0$_{(9)}$ ,7,0$_{(9)}$,15) &$\alpha$ &1.135&88.4\%&1.230&82.0\%&1.639&82.4\% \\
 &$\lambda_1$ &0.492&89.2\%&0.526&84.8\%&0.780&88.6\%\\
 & $\lambda_2$ &0.703&82.8\%&0.740&80.2\%&1.378&88.0\%\\
 \midrule
 k=20,R=(0$_{(18)}$,7,15) & $\alpha$ &1.157&86.6\%&1.239&84.4\%&1.702&82.4\%\\
 & $\lambda_1$ &0.509&89.6\%&0.527&85.2\%&0.800&88.8\%\\
 & $\lambda_2$ &0.694&84.2\%&0.704&82.8\%&1.416&86.8\%\\
 \bottomrule
\end{tabular} 
}
\end{center}
\end{table}

Now comparing the results between the ordered inference and the unordered inference, it is observed that the estimators obtained using
ordered information are slightly better than the unordered ones both the in terms of biases and MSEs.  Moreover, the confidence 
intervals and the credible intervals based on the ordered inference are slightly smaller than the unordered ones.  Hence, if we have 
some knowledge about the ordering on the scale parameters, it is better to use that information.

\subsection{\sc Data  Analysis}

In this section we present the analysis of real data sets to show how the different methods work in practice.  The data represent the strength measured in GPA for single carbon fibers and can be obtained from Kundu and Gupta (2006).  Single fibers were tested under tension at different gauge lengths.  Data set 1 are measurements on single fiber of gauge length 20 mm and Data set 2 are obtained from single fiber of gauge length 10 mm.  They are presented below for an easy reference:\\
Data  set 1:\\
1.312,1.314,1.479,1.552,1.700,1.803,1.861,1.865,1.944,1.958,1.966,1.997,2.006,2.021,
2.027,2.055,
2.063,2.098,2.140,2.179,2.224,2.240,2.253,2.270,2.272,2.274,2.301,2.301,2.359,2.382,2.382,2.426,
2.434,2.435,2.478,2.490,2.511,2.514,2.535,2.554,2.566,2.570,2.586,2.629,2.633,2.642,2.648,2.684,
2.697,2.726,2.770,2.773,2.800,2.809,2.818,2.821,2.848,2.880,2.954,3.012,3.067,3.084,3.090,3.096,
3.128,3.233,3.433,3.585,3.585.\\
Data set 2:\\1.901,2.132,2.203,2.228,2.257,2.350,2.361,2.396,2.397,2.445,2.454,2.474,2.518,2.522,2.525,2.532,
2.575,2.614,2.616,2.618,2.624,2.659,2.675,2.738,2.740,2.856,2.917,2.928,2.937,2.937,2.977,2.996,
3.030,3.125,3.139,3.145,3.220,3.223,3.235,3.243,3.264,3.272,3.294,3.332,3.346,3.377,3.408,3.435,
3.493,3.501,3.537,3.554,3.562,3.628,3.852,3.871,3.886,3.971,4.024,4.027,4.225,4.395,
5.020.

Kundu and Gupta (2006) subtract 0.75 from both the data sets and fit Weibull distributions for both the modified data sets separately.  The MLEs of shape parameters and scale parameters and Kolmogorov-Smirnov (K-S) distances between empirical distribution functions and fitted distribution functions along with $p$ values are provided in Table \ref{data-test1} for both the data sets.  The K-S distances and $p$ values clearly indicate both the data sets after subtracting 0.75 are well fitted by Weibull distributions.

Along with the classical K-S test we also compute the Bayesian counterpart $p_b(data)$, the posterior predictive $p$ value based on the K-S discrepancies, see for example Gelman et al. (1996).   We compute the $p_b(data)$ based on 
the non-informative prior and the results are reported in Table \ref{data-test1-bayesian}.

\begin{table}[h]
\caption{MLEs of parameters, K-S distance and $p$ values for the fitted Weibull models to modified Data set 1 and 2}
\label{data-test1}
\begin{center}
\begin{tabular}{lllll}
\toprule
\multicolumn{1}{c}{data } &\multicolumn{2}{c}{MLE from complete sample} & \multicolumn{1}{c}{K-S distance} & \multicolumn{1}{c}{p value} \\
&\multicolumn{1}{c}{shape parameter }&\multicolumn{1}{c}{scale parameter } & &\\
\midrule
Data set1 & $\alpha= 3.843$& $\lambda=0.088$& 0.046& 0.998 \\
 \midrule
Data set 2 & $\alpha =3.909 $& $\lambda=0.025  $ & 0.079 & 0.815\\  
\bottomrule
\end{tabular} 
\end{center}
\end{table}

\begin{table}[h]
\caption{Bayes estimates of parameters, expected posterior K-S and $p$ values for the fitted Weibull models to modified Data set 1 and 2}
\label{data-test1-bayesian}
\begin{center}
\begin{tabular}{lllll}
\toprule
\multicolumn{1}{c}{data } &\multicolumn{2}{c}{Bayes estimate from complete sample} & \multicolumn{1}{c}{expected posterior } & \multicolumn{1}{c}{$p_b(data)$} \\
&\multicolumn{1}{c}{shape parameter }&\multicolumn{1}{c}{scale parameter } &\multicolumn{1}{l}{K-S distance} &\\
\midrule
Data set1 & $\alpha= 3.930$& $\lambda=0.082$& 0.061& 0.877 \\
 \midrule
Data set 2 & $\alpha =4.051 $& $\lambda=0.022  $ & 0.090 & 0.703\\  
\bottomrule
\end{tabular} 
\end{center}
\end{table}

 To test whether two data sets have common shape parameter, we perform the K-S test as well as conventional likelihood ratio test (L-R).  The MLE of the common shape parameter, two scale parameters along with $p$ values  are recorded in Table \ref{data-test2}. 
 Similarly  we compute the posterior predictive $p$ values $p_b(data)$ to test the equality of the shape parameters based on the K-S test.  
Results are provided in Table \ref{data-test2-bayesian}.  We have obtained the expected $p$ value of the LR test based on the posterior 
density and it is 0.41.  Based on these values, it is reasonable to assume that both data sets are from Weibull distributions with equal shape
parameter.

 \begin{table}[h]
\caption{MLEs of parameters,  p value of L-R and K-S test for  common shape parameter on fitted Weibull models to modified Data set 1 and 2}
\label{data-test2}
\begin{center}
\begin{tabular}{llll}
\toprule

\multicolumn{1}{c}{shape parameter } &\multicolumn{1}{c}{scale parameter} &  \multicolumn{1}{c}{p value(L-R test)}& \multicolumn{1}{c}{p value(K-S test)} \\
\toprule
 $\alpha= 3.876$ & $\lambda_1=0.0861$   & 0.895 & (data set1) 0.998\\
 & $\lambda_2=0.026$ & & (data set 2) 0.852\\
\bottomrule
\end{tabular} 
\end{center}
\end{table}

\begin{table}[h]
\caption{Bayes estimate of parameters,  expected posterior K-S distance and $p$ values for  common shape parameter on fitted Weibull models to modified Data set 1 and 2}
\label{data-test2-bayesian}
\begin{center}
\begin{tabular}{llll}
\toprule
\multicolumn{1}{c}{shape parameter } &\multicolumn{1}{c}{scale parameter} &\multicolumn{1}{c}{expected posterior K-S distance} & \multicolumn{1}{c}{$p_b(data)$} \\
\toprule
 $\alpha= 4.074$ & $\lambda_1=0.0881$ & (data set1) 0.102  & (data set1) 0.552\\
 & $\lambda_2=0.030$ & (data set 2) 0.134 &(data set 2) 0.351\\
\bottomrule
\end{tabular} 
\end{center}
\end{table}

 For illustrative purpose we have generated a joint progressive type-II censored sample with $k=20$, $R_i=4$ for $i=1,\ldots,19$ and $R_{20}=36$.  The data set is as follows:  
(1.312,1,2) , (1.314,1,2), (1.479,1,1) ,(1.552,1,1), (,1.700,1,3), (1.861,1,1) ,(1.865,1,1), (1.901,0,4), (1.944,1,2), (1.966,1,3), (1.997,1,1), (2.006,1,2), (2.027,1,1), (2.055,1,3), (2.098,1,2), (2.132 ,0,3),(2.140,1,2), (2.179 ,1,2), (2.203,0,3), (2.257,0,14).

First we obtain the MLEs of the unknown parameters without any order restriction and the associate $90\%$ confidence intervals.  The Bayes estimators (BEs) and the associated $90\%$ confidence and credible intervals without any order restriction are obtained based on non-informative priors setting $a_0=0,b_0=0, a_1=0, a_2=0$ and $a=0, b=4$.  We set non-informative prior of $\alpha$ as $GA(0,4)$ so that the posterior density $\pi^*_2(\alpha|data)$ is integrable based on the given data.  The results are presented in Table \ref{data-no-1} and Table \ref{data-no-2}. The corresponding results based on order restriction are presented in Table \ref{data-not-1} and Table \ref{data-not-2}. 

\begin{table}[h]
\caption{The MLEs and BEs of the unknown parameters for the real data set}
\label{data-no-1}
\begin{center}
\begin{tabular}{lll}
\toprule
\multicolumn{1}{c}{Parameter} &\multicolumn{1}{c}{MLE} & \multicolumn{1}{c}{BE} \\
\midrule
$\alpha$ & 4.495 & 3.896\\
$\lambda_1$ &0.071 & 0.098\\
$\lambda_2$ & 0.016 &0.028\\
 \bottomrule
\end{tabular} 
\end{center}
\end{table}

\begin{table}[h]
\caption{The CIs and CRIs of the unknown parameters for the  real data set}
\label{data-no-2}
\begin{center}
\begin{tabular}{lllll}
\toprule
\multicolumn{1}{c}{Parameter} &\multicolumn{2}{c}{ 90\% HPD CRI} & \multicolumn{2}{l}{90\% Bootstrap CI} \\
& \multicolumn{1}{l}{LL} & \multicolumn{1}{l}{UL}
&\multicolumn{1}{l}{LL}  &\multicolumn{1}{l}{UL} \\
\midrule
$\alpha$ & 3.472 & 4.338 &3.461 &6.693 \\ 
$\lambda_1$ & 0.070&0.167 &0.030&0.117  \\  
$\lambda_2$ &0.007 & 0.049&0.0004 & 0.037  \\  
 \bottomrule
\end{tabular} 
\end{center}
\end{table}

\begin{table}[h]
\caption{The MLEs and BEs of the unknown parameters for the  real data set with order restriction}
\label{data-not-1}
\begin{center}
\begin{tabular}{lll}
\toprule
\multicolumn{1}{c}{Parameter} &\multicolumn{1}{c}{MLE} & \multicolumn{1}{c}{BE} \\
\midrule
$\alpha$ & 4.495 & 3.728\\
$\lambda_1$ & 0.071 & 0.088\\
$\lambda_2$ &0.016 &0.022\\
 \bottomrule
\end{tabular} 
\end{center}
\end{table}

\begin{table}[h]
\caption{The CIs and CRIs of the unknown parameters for the  real data set with order restriction}
\label{data-not-2}
\begin{center}
\begin{tabular}{lllll}
\toprule
\multicolumn{1}{c}{Parameter} &\multicolumn{2}{c}{ 90\% HPD CRI} & \multicolumn{2}{l}{90\% Bootstrap CI} \\
& \multicolumn{1}{l}{LL} & \multicolumn{1}{l}{UL}
&\multicolumn{1}{l}{LL}  &\multicolumn{1}{l}{UL} \\
\midrule
$\alpha$ & 3.442 & 4.360 & 3.461 & 6.774\\ 
$\lambda_1$ & 0.078 & 0.112 & 0.028 & 0.117  \\  
$\lambda_2$ & 0.007 & 0.033 &  0.004 & 0.038  \\  
 \bottomrule
\end{tabular} 
\end{center}
\end{table}

Based on  the joint type-II progressively censored data, among 
non-informative Bayes estimates and MLEs, non-informative Bayes estimates are more close to the estimates based on complete data sets (see Table \ref{data-test2}) 
Also the length of the HPD credible intervals are less than that of the Bootstrap CIs.
   
\section{\sc Conclusions}

In this paper we consider the analysis of joint progressively censored data for two populations.  It is assumed that the lifetime 
of the items of the two populations follow Weibull distributions with the same shape parameter but different scale parameters.  We 
obtain the MLEs of the unknown parameters, and showed that they always exist and they are unique.  We also obtain the Bayes estimate
and the associated credible interval of a function of the unknown parameters based on a fairly general prior distribution both on the
scale and shape parameters.  We further consider the order restricted inference on the unknown parameters.  Based on an extensive 
simulation experiment it is observed that the Bayes estimators with respect to the informative priors perform significantly better
than the corresponding Bayes estimators based on non-informative priors for point estimation in terms of average bias and mean squared errors.  Also non-informative prior based estimators perform slightly better than MLEs
in terms of average bias and mean squared errors.   The credible intervals based on an informative prior perform better 
than the other two methods in terms of average lengths and coverage percentage.  Also it is observed that if the ordering 
information on the scale parameters is available, it is better to use it.

It may be mentioned that in this paper we have mainly considered 
two populations.  It will be interesting to extend the results  for more than two populations also.  Su (2013) developed joint progressive censoring scheme for multiple  populations.  For multiple Weibull populations with common shape parameter, we can assume multivariate gamma-Dirichlet prior for scale parameters and a log-concave prior for common shape parameter as described in this article.  More work is needed along that 
direction.

\noindent {\sc Acknowledgements:} The authors would like to thank the Editor-in-Chief Prof. D. Dey for providing 
several important suggestions which have helped to improve the manuscript significantly.

\section*{\sc Appendix A: Proof of Theorem 1}

To prove Theorem 1, first we show the following result.  Suppose $a_j \ge 0$, for $j = 1, \ldots, k$, $\ds g(\alpha) = \sum_{j=1}^k
a_j t_j^{\alpha}$, then -$\ds \ln(g(\alpha))$ is a concave function.  Note that 
$$
g'(\alpha) = \sum_{j=1}^k a_j t_j^{\alpha} \ln t_j \ \ \  \hbox{and} \ \ \ \
g''(\alpha) = \sum_{j=1}^k a_j t_j^{\alpha} (\ln t_j)^2.
$$
Moreover,
$$
\left (\sum_{j=1}^k a_j t_j^{\alpha} (\ln t_j)^2 \right ) \left ( \sum_{j=1}^m a_j t_j^{\alpha} \right ) - \left (\sum_{j=1}^m a_j t_j^{\alpha} \ln t_j
\right )^2 = \sum_{1 \le i < j \le k} a_i a_j (\ln t_i - \ln t_j)^2 \ge 0.
$$
Therefore,
$$
-\frac{d^2 \ln g(\alpha)}{d \alpha^2}  = - \frac{g''(\alpha) g(\alpha) - (g'(\alpha))^2}{(g(\alpha))^2} \le 0.
$$
Now from the above observation, it immediately follows that $p(\alpha)$ is a concave function.  The result follows by observing the 
fact that $p(\alpha) \rightarrow -\infty$, as $\alpha \rightarrow 0$ or $\alpha \rightarrow \infty$.    \qed

\section*{\sc Appendix B: Fisher Information Matrix}

If we denote the matrix ${\bf A}$ as
$$
{\bf A} = - \left [ \matrix{\frac{\partial^2 l}{\partial \alpha^2} & \frac{\partial^2 l}{\partial \alpha \partial \lambda_1}
&  \frac{\partial^2 l}{\partial \alpha \partial \lambda_2}  \cr
\frac{\partial^2 l}{\partial \lambda_1  \partial \alpha} & \frac{\partial^2 l}{\partial \lambda_1^2} & 
\frac{\partial^2 l}{\partial \lambda_1  \partial \lambda_2} \cr
\frac{\partial^2 l}{\partial \lambda_2  \partial \alpha} & \frac{\partial^2 l}{\partial \lambda_2  \partial \lambda_1} &
\frac{\partial^2 l}{\partial \lambda_2^2} \cr} \right ] = \left [ \matrix{a_{11} & a_{12} & a_{13}  \cr a_{21} & a_{22} & a_{23} 
\cr a_{31} & a_{32} & a_{33} \cr} \right ],
$$
then
\beanno
a_{11} & = & \frac{k}{\alpha^2} + \lambda_1 \left [ \sum_{j=1}^k s_j t_j^{\alpha} (\ln t_j)^2 + \sum_{C_1} t_j^{\alpha} (\ln t_j)^2 
\right ] + \lambda_2 \left [ \sum_{j=1}^k w_j t_j^{\alpha} (\ln t_j)^2 + \sum_{C_2} t_j^{\alpha} (\ln t_j)^2 
\right ]  \\
a_{12} & = & a_{21} =  \left [ \sum_{j=1}^k s_j t_j^{\alpha} \ln t_j + \sum_{C_1} t_j^{\alpha} \ln t_j \right ]   \\
a_{13} & = & a_{31} =  \left [ \sum_{j=1}^k w_j t_j^{\alpha} \ln t_j + \sum_{C_2} t_j^{\alpha} \ln t_j \right ]  \\
a_{22} & = & \frac{k_1}{\lambda_1^2}, \ \ a_{33} = \frac{k_1}{\lambda_1^2}, \ \ a_{23} = a_{32} = 0.
\eeanno

\end{document}